\documentclass{PoS}

\title{CSS galaxy embedded within the core of a bright X-ray cluster}

\ShortTitle{CSS galaxy embedded within the core of a bright X-ray cluster}

\author{\speaker{Maciej Ceglowski}%
         \\
        Torun Centre for Astronomy, Gagarina 11, 87-100 Torun, Poland\\
        E-mail: \email{ceglowski@astri.uni.torun.pl}}

\author{Magdalena Kunert-Bajraszewska\\
        Torun Centre for Astronomy, Gagarina 11, 87-100 Torun, Poland\\
        E-mail: \email{magda@astro.uni.torun.pl}}
\author{A. Siemiginowska\\
       Harvard-Smithsonian Center for Astrophysics, 60 Garden Street, Cambridge, MA 02138, USA \\
}
        
 \author{ A. Labiano\\
        Centro de Astrobiologia (CSIC-INTA), Carretera de Ajalvir km. 4, 28850 Torrejon de Ardoz, Madrid, Spain.\\
} 
        
  \author{M. Guainazzi\\
       European Space Astronomy Centre of ESA, PO Box 78, Villanueva de la
Ca\~nada, 28691, Madrid, Spain \\
        
        }      
        
\abstract{We discovered an X-ray cluster in a recent
pointed {\it Chandra} observation of
the radio-loud compact-steep-spectrum source
1321+045 at the redshift of 0.263. 1321+045 is  part of larger survey which aims to study 
the X-rays properties of weak compact radio sources. Compact 
radio sources are young objects at the beginning of their evolution and if embedded in
an X-ray cluster offer unique opportunities to study the cluster heating
process.}

\FullConference{11th European VLBI Network Symposium \& Users Meeting\\
                 9-12 October 2012\\
                 Bordeaux (France)}

\begin{document}

\section{Introduction}
Clusters of galaxies are the largest gravitationally bound objects in the Universe. The
space between the galaxies is filled with a hot 
plasma that emits X rays. High-resolution images of
nearby cluster cores taken by the {\it Chandra} X-ray Observatory show that 
 large amounts of energy are injected into the surrounding medium by
powerful outbursts of active galactic nuclei (AGN)
lying at the hearts of galaxy clusters. The combination of
high-resolution X-ray and radio imaging provides a direct measure of 
this energy, which is in most cases
sufficient to prevent cooling of the intra-cluster medium and the substantial
growth at late times of giant elliptical (gE) and cD
galaxies \cite{mac09}.

Most X-ray clusters are found around radio-loud
active galaxies with large-scale radio structures
and the majority of them is classified as FR\,Is \cite{fr}. 
The FR\,I radio sources are old (> $10^{7} $ years) and their long-term 
interaction with the cluster environment 
imprinted a rich variety of structures into the X-ray morphology, 
such as bubbles, shock fronts and ripples \cite{mac09,sul12}. 
Gigahertz Peaked Spectrum (GPS)
and Compact Steep Spectrum (CSS) radio sources are
young (< $10^{5}$ years) and have not developed large-scale
radio structures (typical size <20 kpc\footnote{We
use the following cosmology: 
${\rm
H_0}$=71${\rm\,km\,s^{-1}\,Mpc^{-1}}$, $\Omega_{M}$=0.27,
$\Omega_{\Lambda}$=0.73.} 
\cite{odea98}); and they are believed to be in the beginning
of their evolution \cite{rea96}. These objects, if found in clusters,
can potentially test the cluster heating process and the
significance of the AGN in the evolution of the cluster.
The first X-ray cluster known to host a compact steep spectrum radio
source was 3C\,186 discovered by Siemiginowska et al. \cite{siem05,siem10}. The
X-ray morphology indicates that the cluster is well formed and     
has a cool core with a short central cooling time. The radio source can
potentially supply
the energy required to stabilize the cluster core against catastrophic
cooling, as it expands into the cluster medium. 

The 1321+045 radio source presented here is the second compact steep spectrum radio source
known to be associated with a large X-ray cluster. It has a
radio morphology  different from  that of  3C\,186 (FR\,I vs FR\,II-like) and
it is much less luminous at radio waves. These two cases indicate that the 
radio source properties may be different even though they are located in a
similar cluster environment. They provide a great opportunity to investigate 
the radio source evolution and interaction with the cluster environment.

\section{Sample and observations}
\begin{figure*}
\centering
\includegraphics[width=9cm, height=8cm]{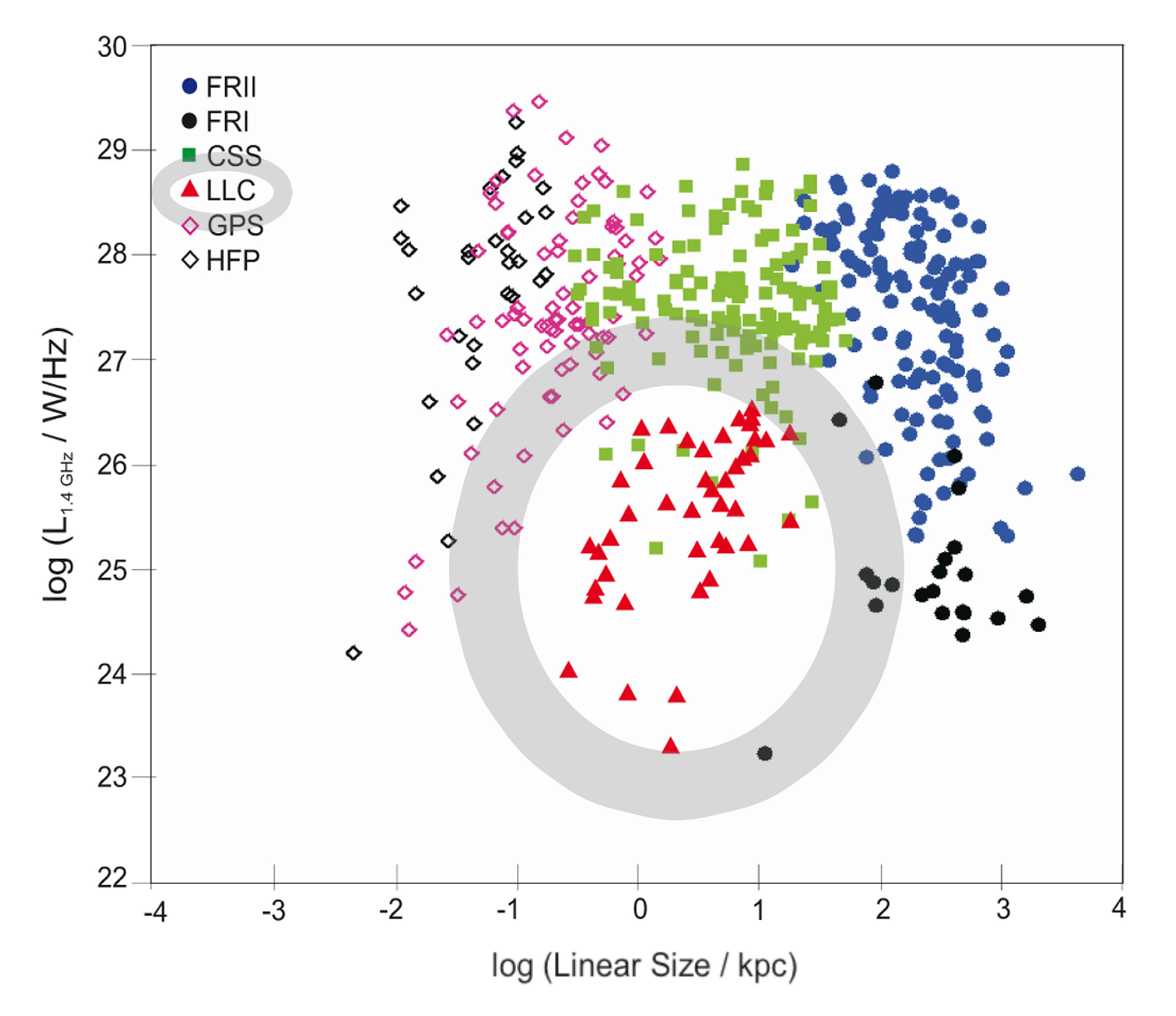}
\caption{Evolutionary diagram of AGN taken from \cite{kunert10a}.
LLC objects occupy a poorly sampled locus (grey circle) in
the diagram.}
\label{fig1}
\end{figure*}

1321+045 is one of the seven radio sources observed with {\it Chandra}
ACIS-S3 using 1/8 subarray and standard pointings. The observations of the
whole sample were performed in February 2011 and June 2012, with
exposure times of $\sim$9.5\,ks for each source (see Table \ref{tab1}). 
The target sources belong to the sample  
of Low Luminosity Compact (LLC) objects
and their radio and optical properties were analyzed in \cite{kunert10a} and
\cite{kunert10b}, respectively.  
The main criterion for selecting  LLC objects  is the radio luminosity threshold. 
All sources fall below $L_{1.4\,GHz} < 10^{26} {\rm W~Hz}^{-1}$ and occupy a poorly-studied 
locus in the radio power versus linear size diagram (Fig. \ref{fig1})
\cite{kunert10a}. 
We believe that it means that the majority of the LLC objects are short-lived 
objects, at least in the current phase of evolution.
They can go up and down on the main evolutionary scheme, restarting their 
activity many times. 
Most probably one of the crucial stages for AGN is when they leave their host galaxy. 
At this stage, the interaction of the radio jets with the interstellar medium (ISM) can
disrupt the jet and change the morphology and luminosity of the
source, which is likely to affect its evolution. 

Our {\it Chandra} targets were selected based on their radio morphology
to represent a different evolutionary state of the source growth, so they have either a
weak or undetected radio core and strong lobes, or breaking up radio lobes with a bright radio core.
All sources have been resolved by MERLIN at 1.6\,GHz and have a CSO (compact symmetric object)-type or complex radio morphology.
Their linear radio sizes are in the range 2--16\,kpc \cite{kunert10a}.

There is a limited number of known bright compact AGNs with X-ray data available
\cite{gan04,vink06,siem08,messi09,ten09}, however  
no X-ray data have been obtained for low power CSS sources so far. It is thus expected that our
recent {\it Chandra} observations will establish the X-ray properties of these sources for the first 
time. Our project aims at answering several important questions:
(1) determine the locus of the low radio power CSS sources in the radio to X-ray 
luminosity parameter space; (2) study the effect of the environment on
the deceleration of the jet \cite{gan04}; (3) analyze the correlation
between the absorption column and the size of the radio source using the new 
data together with archival X-ray observations \cite{ten09};
(4) determine if there is any diffuse X-ray emission exceeding nuclear scales in these sources.

\begin{table} 
\caption{Basic parameters of the 7 LLC objects observed with Chandra.}
\begin{tabular}{l l c c c c c}
\hline
Source name& 
\multicolumn{1}{c}{\it z}& 
Obs ID& ${\rm Counts_{0.5-7\,keV}}$& ${\rm
log(L_{5\,GHz})}$& $\Gamma$& ${\rm log(L_{2-10\,keV})}$ \\
\multicolumn{1}{c}{(1)}&
\multicolumn{1}{c}{(2)}&
(3)&(4)&(5)&(6)&(7)\\   
\hline
0810+077& 0.112          &12716 &119 (11)     &41.4& $0.64^{+0.15}_{-0.15}$ &42.9 \\
0907+049& $0.640^{phot}$ &12717 &3$^{a}$   &42.6& 1.7$^{b}$ &$42.9^{c} $ \\
0942+355& 0.208          &12714 &103 (10)      &41.4& $1.6^{+0.24}_{-0.16}$ &42.9 \\
1321+045& 0.263          &12715 &53 (7)      &41.6& $2.35^{+0.39}_{-0.36}$ &42.3 \\
1542-390& 0.553          &12718 &0$^{a}$   &42.4& 1.7$^{b}$ &$42.7^{c} $ \\
1558+536& 0.179          &12719 &9 (3)        &41.4& 1.7$^{b}$ &41.7 \\
1624+049& $0.040^{phot}$ &12720 &4$^{a}$   &40.0& 1.7$^{b}$ &$40.0^{c} $	\\
\hline	
\label{tab1}
\end{tabular}
\hspace{0.5cm}\\
{\footnotesize
(1) Source name; (2) Redshift, 'phot' indicates a photometric redshift; (3) Observation ID; (4) Counts
number, only upper limits are given for sources with a $^{a}$ symbol, numbers in parentheses indicate the errors calculated as $\sqrt{{\rm counts}}$; 
(5) Luminosity at 5\,GHz taken from \cite{kunert10a} in ${\rm erg~s^{-1}}$; (6)
Photon index, $^{b}$ means $\Gamma$=1.7 was assumed for the flux calculation; (7) Luminosity at 2--10\,keV 
in ${\rm erg~s^{-1}}$, $^{c}$ means only an upper limit is available.}
\end{table}

\section{Preliminary results}
We have detected all but one target source (1542+390) in the X-rays. We used the software package Sherpa
to fit the spectra, assuming an absorbed power-law in the 0.5--7\,keV energy
range. The Galactic absorption was kept frozen during the fitting. 
The second absorption component was
assumed to be intrinsic to the quasar and located at the redshift
of the source. The model was applied to six detected sources. However, since
0907+049, 1558+536 and 1624+049 did not have enough counts
to produce a reasonable fit, we assumed a photon index value
$\Gamma = 1.7$ (see Table 1) to derive the 
flux of these three sources. We have placed our  low-power compact
sources on the X-ray vs radio luminosity plot, together with known strong CSS
and GPS objects and large scale FR\,Is and FR\,IIs. 
Their locus (within the FR Is that are weaker in the X-rays) is consistent with
the established X-ray - radio luminosity correlation for AGN. A detailed study of the whole sample
will be presented in a forthcoming paper.

\begin{figure*}
\centering
\includegraphics[width=0.9\textwidth,height=0.3\textheight]{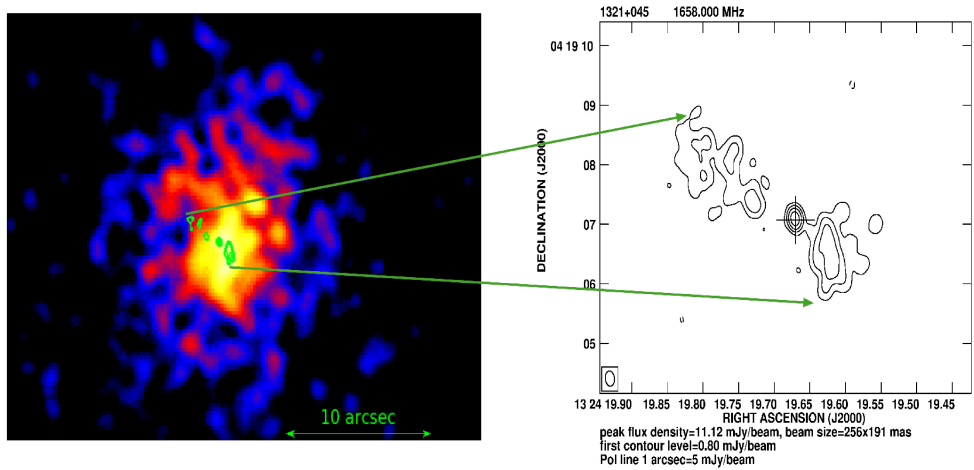}
\caption{(Left) 
{\it Chandra} ACIS-S X-ray image of 1321+045 in the  0.5--7\,keV energy range. The
  original image was smoothed with a Gaussian function
  of width 2 arcsecond. A 10 arcsec  
  scale is marked in the image.  The X-ray color image is overlayed with
  the green radio contours from the MERLIN 1.6\,GHz image. (Right)
  MERLIN radio image of 1321+045 at 1.6\,GHz \cite{kunert10a}. Contour
levels increase by the factor 2 and the first
contour level corresponds to 0.80 mJy/beam (3$\sigma$). The position of the optical counterpart
is marked with a cross and is taken from the Sloan Digital Sky Survey
(SDSS).
}
\label{fig2}
\end{figure*}

Diffuse X-ray emission from a cluster of galaxies was only found
around one of the sources in our sample, 1321+045 (see Fig. \ref{fig2}, left panel). The position of the central
bright component visible in the 1.6\,GHz MERLIN image of this source is well correlated with
the position of the optical counterpart suggesting this component is the radio core,
on opposite sides of which there is diffuse emission from  two radio
lobes (Fig. 2, right panel). This rather symmetric structure has a total length of 17\,kpc. The {\it Chandra} observations provided an 
X-ray luminosity of $10^{42.3}$ erg/s (Table \ref{tab1}), a cluster temperature kT = $4.4^{+0.5}_{-0.3}$\,keV   
and an X-ray morphology typical of other massive, relaxed
clusters. We have obtained the cluster  
temperature and density profiles which suggest that the cluster has a
cooling core. There is no evidence for the presence of any ripples or discontinuities in the
X-ray image of the cluster. Instead, there is a rather uniform emission without indications of
interaction with the radio jets. The detailed analysis of this cluster
is presented in a separate paper \cite{kun2013}.

% {\it Chandra} observations of GPS/CSS sources brought surprising
% results so far, among which the  discovery of X-ray jets, of an X-ray cluster
% at z=1 associated with the radio-loud CSS source 3C186 \cite{siem05,siem10},
% and of thermal diffuse X-ray emission
% associated with an emission line region in 3C305 \cite{messi09}. 1321+045 is
% the first low-power CSS object associated with an X-ray cluster.

{\it Chandra} observations of GPS/CSS sources brought surprising
results so far. One of which was the detection of X-ray emission in the jets of such sources 
\cite{siem08}. 
% and an X-ray cluster at z=1 associated with the radio-loud CSS source 3C186
Additionally, an X-ray cluster at z=1 associated with the radio-loud CSS source 3C186 was unveiled  
\cite{siem05,siem10}.
Another extraordinary example is the discovery of thermal diffuse X-ray emission
associated with an emission line region in 3C305 \cite{messi09}. Finally, the most recent
intriguing outcome from Chandra observations is the discovery of the first low-power CSS object 
(1321+045) associated with an X-ray cluster.

% discovery of the 1321+045 the first 
% low-power CSS object associated with an X-ray cluster.

.

\bigskip
\noindent
{\small
{\bf Acknowledgements}\\
\noindent
MERLIN is a UK National Facility operated by the University of Manchester on
behalf of STFC. This research has made use of data obtained by the {\it Chandra} X-ray
Observatory and of the {\it Chandra} X-ray Center (CXC) software packages CIAO, ChIPS, and Sherpa.
This research is funded in part by NASA contract NAS8-39073. Partial support for this work was provided  
by the {\it Chandra} grants GO1-12124X.

\end{document}